# Causality issue in the heat balance method for calculating the design heating and cooling load

Author name:           Christian Ghiaus
Author affiliation:    INSA-Lyon, CETHIL UMR5008, F-69621 Villeurbanne, France

Contact information for the corresponding author
Mailing address        CETHIL
                       INSA Lyon
                       9 rue de la physique
                       69621 Villeurbanne
                       France
E-mail address:        christian.ghiaus@insa-lyon.fr

**Abstract**
Heating and cooling load calculation based on dynamic models is widely used in simulation software and it is the method recommended by ASHRAE and CEN. The principle is to make the heat balance for the air volume of a room space considered at uniform temperature and to calculate from this equation the load, i.e. the power needed to obtain the required indoor temperature. The problem is that, by doing so, the physical causality is not respected. If the model is approximated by a piecewise linear dynamical system, this procedure results in an improper transfer function. In order to point out this problem, a method to obtain state space and transfer function models from thermal networks is introduced. Then, the transfer function representation is employed to show that changing the physical causality results in an improper transfer function. The practical consequence is that when the space temperature has a step variation, the calculated load tends to infinity if the simulation time step tends to zero. The issue of causality may be a problem in equation-based simulation software, such as MODELICA, in which the equations do not represent causal relations: a wrong choice of the causality in a balance equation may result in improper transfer functions.

**Keywords:** thermal network, state space model, proper and improper transfer function, linear algebra

# 1 Introduction

Estimating the heating or cooling load means to find out what is the power needed by the building in order to maintain the indoor temperature at a desired value when the heat gains and losses (due to weather, ventilation rates, occupancy, usage of electrical appliances, etc.) vary. Thermal load calculation is one of the most important tasks of building design. It is used to improve the architectural design by estimating the energy performance of buildings, to size the heating, cooling and ventilation systems, to conceive the control strategies for lowering the peak demand and the energy consumption, to analyze the life cycle cost. Thermal simulation of buildings by employing the heating and cooling load is used to check if a building design complies with standards and regulations.

It is generally accepted that the heat balance method is fundamental for the thermal load calculation. Other procedures are seen as variants or simplifications of this method. Therefore, heat balance method is used by simulation software such as TRNSYS (TRNSYS 2007), EnergyPlus (EnergyPlus 2012), ESP-r (Clarke 2001), CLIMA (Skeiker 2010), etc. It is the recommended method of ASHRAE (Bellenger et al. 2001) and the basis of the CEN standard for calculation of the design heat load (CEN 2003) as well as of other standards related to thermal performances of buildings (e.g. CEN 2007a, CEN 2007b).

The principle of the method is to do the heat balance on the air volume of the room. This implies calculating the conductive, convective and radiative heat balance for each surface of the room and the convective heat balance for the room air (e.g. Wang and Chen 2000, Bellenger et al. 2001, Hernandez *et al.* 2003, Xu and Wang 2008, Li *et al.* 2009).

The result of the heat balance is an input-output model. A dynamical input-output system is causal when its output depends only on past inputs and noncausal if its output depends also on future inputs. All physical systems need to be causal (Romagnoli and Palazoglu 2012). An acausal system is not physically realizable because it requires prediction (Morari and Zafiriou 1989).

When used for calculating the indoor temperature, the heat balance method is a direct method (i.e. it respects the physical causality): the output, i.e. the indoor temperature, depends on more inputs, including the thermal load. However, when it is used to estimate the load, the heat balance method, in its actual form, results in an acausal system: the output of the physical system, i.e. the indoor temperature, becomes an input of the algorithm for calculating the thermal load. The practical consequence of the acausality is that, if the indoor room temperature has a step variation, the calculated load tends to infinity when the time step for calculation tends to zero. Mathematically, if the thermal model is linear (or piece-wise linearized for one or more time steps during the simulation), it is characterized by an improper transfer function which grows unbounded as the frequency approaches infinity. Practically, when the indoor temperature varies, the value of the calculated thermal load will depend on the value of the time step used in simulation. The physical explanation is rather intuitive: an infinite power is needed to change the indoor temperature instantaneously. Therefore, the sizing of the HVAC system, which is determined by the calculated load, will depend on the time step used in simulation. Practitioners noticed this problem and introduced load limitations in the heat balance method (e.g. CEN 2007b).

The way in which the heat balance method for load calculation is usually presented makes it difficult to notice that the model used for estimating the load results in an improper transfer function that does not respect the causality. However, if the whole system of equations is written in state space form and then transformed into a set of transfer functions, it becomes evident that some transfer functions are improper, which reflects their acausality.

## 2 Principle of the heat balance method

The most important assumption of the heat balance (HB) method is that the temperature of the air in the thermal zone is homogeneous. The other important assumptions are that the surfaces (walls, windows, floors, ceilings, etc.) have uniform temperature, uniform long-wave and short-wave irradiation and diffuse radiation (Bellenger et al. 2001). The heat conduction is usually treated in one dimension but two or three dimensions may be handled. Based on these assumptions, heat balance is done for each wall for outside face, nodes in the solid walls and inside face. Finally, the convective heat balance is done for the room air (Figure 1):

$$m_a c_a \frac{d\theta_a}{dt} = q_{ci} + q_v + \dot{Q}_g + \dot{Q}_{HVAC} \qquad (1)$$

where:

$\theta_a$        is the room air temperature considered homogeneous [°C or K];

$m_a c_a \dfrac{d\theta_a}{dt}$    - rate of increase of heat stored in the room air volume of mass $m_a$ and specific heat $c_a$ [W];

$q_{ci} = \sum_i S_i h_i (\theta_{ci} - \theta_a)$ - convective heat flow from surface $S_i$ having the surface temperature $\theta_{ci}$ and the inside heat convection coefficient $h_i$ [W];

$q_v = \dot{m}_v c_a (T_o - \theta_a)$ - heat transfer rate due to mass flow $\dot{m}_v$ of outdoor air at temperature $T_o$ introduced into the space by infiltration and ventilation [W];

$\dot{Q}_g$         - convective heat flow gained from the internal loads [W];

$\dot{Q}_{HVAC}$     - heat flow to/from the HVAC system [W].

If the thermal capacity of the air in the zone is neglected, then equation (1) becomes:

$$q_{ci} + q_v + \dot{Q}_g + \dot{Q}_{HVAC} = 0 \qquad (2)$$

In the heat balance method, the load, $\dot{Q}_{HVAC}$, is determined from the inside air heat balance. If the heat capacity of the room air is not neglected, then

$$\dot{Q}_{HVAC} = m_a c_a \frac{d\theta_a}{dt} - q_{ci} - q_v - \dot{Q}_g \qquad (3)$$

If the heat capacity of the air is neglected, then

$$\dot{Q}_{HVAC} = -q_{ci} - q_v - \dot{Q}_g \qquad (4)$$

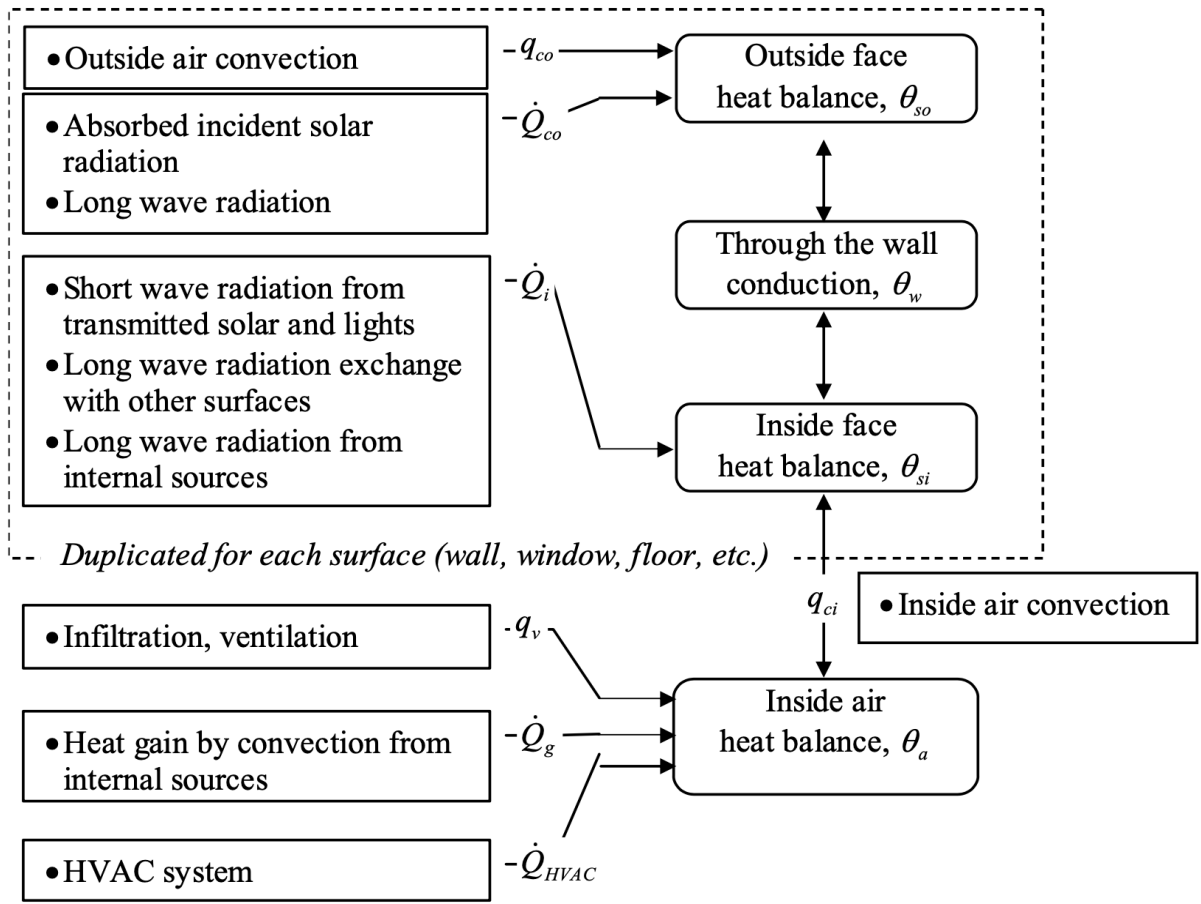

*Figure 1 Heat balance method (adapted after Bellenger et al. 2001)*

In equations (3) and (4) the room air temperature, $\theta_a(t)$, is considered known because it is the desired temperature of the indoor air.

Finding indoor air temperature $\theta_a$ from equations (1) or (2) is a simulation (or direct) problem. Finding the load $\dot{Q}_{HVAC}$ from equation (1) or (2) is a control (or inverse) problem. If the model for heat transfer is linear (or piecewise linear), than equations (3) or (4) result in an improper transfer function. In order to demonstrate this, we will first obtain a state space formulation of the thermal model. Then, from the state space form we will obtain the transfer functions and we will demonstrate that equations (3) or (4) end in an improper transfer function. A simple numerical example, which does not reduce the generality of the problem, will be used to illustrate these steps.

## 3   Differential algebraic equation representation of the thermal model

The thermal model of the room is described by a system of differential algebraic equations.  In order to obtain this system, an algorithm is proposed which is inspired by Strang (1986, 2009), who introduced it for systems in equilibrium, and by the Modified Nodal Analysis method, frequently used in electrical circuits modeling and analysis. As compared to electrical circuits, the thermal circuits do not have an element equivalent to inductance. Thermal circuits may be arranged in branches, containing thermal resistances and temperature sources, and nodes, to which thermal capacities and flow sources are connected.

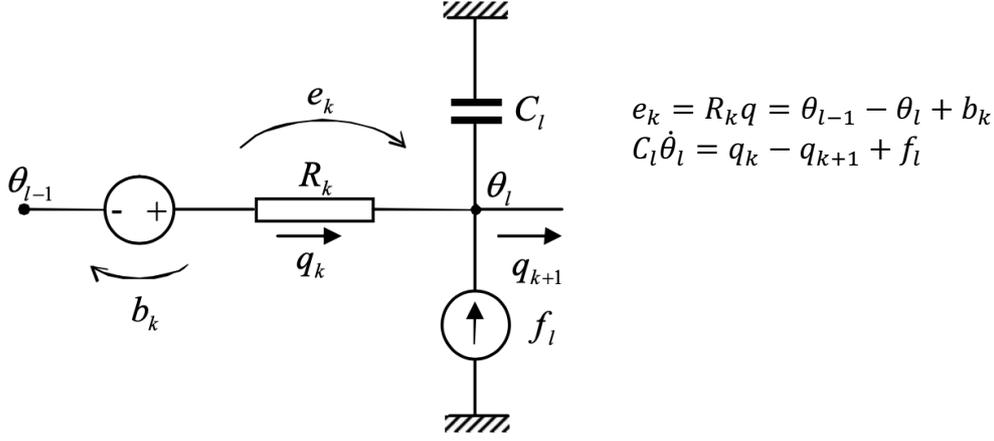

*Figure 2 Typical branch and node in thermal networks*

The temperature difference over a resistance $R_k$ is

$$e_k = \theta_{l-1} + b_k - \theta_l \qquad (5)$$

where $\theta_{l-1}$ and $\theta_l$ are the temperatures in the nodes connected by the branch $k$, and $b_k$ is the temperature source on the branch $k$. Writing equation (5) for all branches, we obtain:

$$\mathbf{e} = -\mathbf{A}\boldsymbol{\theta} + \mathbf{b} \qquad (6)$$

where:

$\mathbf{e} = \begin{bmatrix} e_1 & ... e_k ... & e_m \end{bmatrix}^T$ is the vector of temperature drops over the thermal resistances,

$\boldsymbol{\theta} = \begin{bmatrix} \theta_1 & ... \theta_l ... & \theta_n \end{bmatrix}^T$ is the vector of temperature in the nodes,

$\mathbf{b} = \begin{bmatrix} b_1 & ... b_k ... & b_m \end{bmatrix}^T$ is the vector of temperature sources on the branches.

The matrix $-\mathbf{A}$ is a difference operator for the temperatures while the matrix $\mathbf{A}$ is the incidence matrix of the network representing the thermal circuit. The rows of the matrix $\mathbf{A}$ correspond to the branches containing the resistances $R_k$ (i.e. the edges of the graph) and the columns correspond to the nodes representing the temperatures $\theta_l$ (i.e. the vertices of the graph). The elements of the incidence matrix $\mathbf{A}$ have the value:

$$a_{kl} = \begin{cases} 0, & \text{if } R_k \text{ not connected to node } \theta_l \\ +1, & \text{if flow passing through } R_k \text{ enters in node } \theta_l \\ -1, & \text{if flow passing through } R_k \text{ exists from node } \theta_l \end{cases} \qquad (7)$$

The heat flux in each branch is

$$q_k = R_k^{-1} e_k \tag{8}$$

Expressed for all fluxes in matrix form, equation (8) gives

$$\mathbf{q} = \mathbf{G}\,\mathbf{e} \tag{9}$$

where:

$\mathbf{q} = \begin{bmatrix} q_1 & \ldots q_k \ldots & q_m \end{bmatrix}^T$ is the vector of heat rates in the branches and

$\mathbf{G} = \begin{bmatrix} R_1^{-1} & 0 & 0 \\ 0 & \ddots & 0 \\ 0 & 0 & R_m^{-1} \end{bmatrix}$ is a diagonal matrix of thermal conductivities.

The balance of heat rates in a node $\theta_l$ states that the variation in time of the energy accumulated in the thermal capacity, $C_l \dot{\theta}_l$, is equal to the algebraic sum of heat rates entering the node, $\sum_l q_l$ and the heat flow sources $f_l$ connected to the node:

$$C_l \dot{\theta}_l = \sum_l q_l + f_l \tag{10}$$

Writing the balance equation for all nodes in matrix form gives:

$$\mathbf{C}\dot{\boldsymbol{\theta}} = \mathbf{A}^T \mathbf{q} + \mathbf{f} \tag{11}$$

where:

$\mathbf{C} = \begin{bmatrix} C_1 & 0 & 0 \\ 0 & \ddots & 0 \\ 0 & 0 & C_n \end{bmatrix}$ is a diagonal matrix of thermal capacities;

$\mathbf{A}^T$, the transpose of the incidence matrix, is a matrix operator which makes the algebraic sum of heat transfer rates $\mathbf{q}$;

$\mathbf{f} = \begin{bmatrix} f_1 & \ldots f_l \ldots & f_n \end{bmatrix}^T$ is the vector of heat rate sources in each temperature node.

By substituting $\mathbf{e} = -\mathbf{A}\boldsymbol{\theta} + \mathbf{b}$ from equation (6) into equation (9) and then $\mathbf{q} = \mathbf{G}\mathbf{e}$ from equation (9) into equation (11), we obtain:

$$\mathbf{C}\dot{\boldsymbol{\theta}} = -\mathbf{A}^T \mathbf{G} \mathbf{A} \boldsymbol{\theta} + \mathbf{A}^T \mathbf{G} \mathbf{b} + \mathbf{f} \tag{12}$$

or

$$\mathbf{C}\dot{\boldsymbol{\theta}} = \mathbf{K}\boldsymbol{\theta} + \mathbf{K}_b \mathbf{b} + \mathbf{f} \tag{13}$$

where we note $\mathbf{K} \equiv -\mathbf{A}^T \mathbf{G} \mathbf{A}$ and $\mathbf{K}_b \equiv \mathbf{A}^T \mathbf{G}$.

Since some elements of the diagonal matrix $\mathbf{C}$ are zero, the relations (12) and (13) represent a system of differential algebraic equations (DAE): the equations corresponding to zero elements on the diagonal of matrix $\mathbf{C}$ are algebraic while the equations corresponding to non-zero elements on the diagonal of matrix $\mathbf{C}$ are differential.

## 4   State space representation of the thermal model

Equation (13) can be written in state space form by eliminating the temperatures corresponding to the nodes without thermal capacity, i.e. the nodes $l$ for which $C_{ll} = 0$. Equation (13) can be written in bloc-matrix form as:

$$
\begin{bmatrix} \mathbf{0} & \mathbf{0} \\ \mathbf{0} & \mathbf{C}_C \end{bmatrix} \begin{bmatrix} \dot{\boldsymbol{\theta}}_0 \\ \dot{\boldsymbol{\theta}}_C \end{bmatrix} = \begin{bmatrix} \mathbf{K}_{11} & \mathbf{K}_{12} \\ \mathbf{K}_{21} & \mathbf{K}_{22} \end{bmatrix} \begin{bmatrix} \boldsymbol{\theta}_0 \\ \boldsymbol{\theta}_C \end{bmatrix} + \begin{bmatrix} \mathbf{K}_{b1} \\ \mathbf{K}_{b2} \end{bmatrix} \mathbf{b} + \begin{bmatrix} \mathbf{I}_{11} & \mathbf{0} \\ \mathbf{0} & \mathbf{I}_{22} \end{bmatrix} \begin{bmatrix} \mathbf{f}_0 \\ \mathbf{f}_C \end{bmatrix} \tag{14}
$$

where:

$\boldsymbol{\theta}_0$ and $\mathbf{f}_0$ correspond to the nodes without thermal capacity;

$\boldsymbol{\theta}_C$ and $\mathbf{f}_C$ correspond to the nodes with thermal capacity;

$\mathbf{C}_C$ is the bloc of the partitioned matrix $\mathbf{C}$ for which the elements are non-zero;

$\mathbf{K}_{11}$, $\mathbf{K}_{12}$, $\mathbf{K}_{21}$, $\mathbf{K}_{22}$ are blocs of the partitioned matrix $\mathbf{K}$ obtained according to the partitioning of the matrix $\mathbf{C}$;

$\mathbf{K}_{b1}$, $\mathbf{K}_{b2}$ are blocs of the partitioned matrix $\mathbf{K}_b$ obtained according to the partitioning of the matrix $\mathbf{C}$;

$\mathbf{I}_{11}$, $\mathbf{I}_{22}$ are identity matrices.

The vector $\boldsymbol{\theta}_0$ needs to be eliminated from equation (14). By multiplying the first row of equation (14) by $-\mathbf{K}_{21}\mathbf{K}_{11}^{-1}$, we obtain:

$$
\begin{aligned}
\begin{bmatrix} \mathbf{0} & \mathbf{0} \\ \mathbf{0} & \mathbf{C}_C \end{bmatrix} \begin{bmatrix} \dot{\boldsymbol{\theta}}_0 \\ \dot{\boldsymbol{\theta}}_C \end{bmatrix} &= \begin{bmatrix} -\mathbf{K}_{21} & -\mathbf{K}_{21}\mathbf{K}_{11}^{-1}\mathbf{K}_{12} \\ \mathbf{K}_{21} & \mathbf{K}_{22} \end{bmatrix} \begin{bmatrix} \boldsymbol{\theta}_0 \\ \boldsymbol{\theta}_C \end{bmatrix} + \\
&\quad \begin{bmatrix} -\mathbf{K}_{21}\mathbf{K}_{11}^{-1}\mathbf{K}_{b1} \\ \mathbf{K}_{b2} \end{bmatrix} \mathbf{b} + \begin{bmatrix} -\mathbf{K}_{21}\mathbf{K}_{11}^{-1} & \mathbf{0} \\ \mathbf{0} & \mathbf{I}_{22} \end{bmatrix} \begin{bmatrix} \mathbf{f}_0 \\ \mathbf{f}_C \end{bmatrix}
\end{aligned} \tag{15}
$$

By replacing the second row with the sum of the two rows, equation (15) becomes:

$$
\begin{aligned}
\begin{bmatrix} \mathbf{0} & \mathbf{C}_C \end{bmatrix} \begin{bmatrix} \dot{\boldsymbol{\theta}}_0 \\ \dot{\boldsymbol{\theta}}_C \end{bmatrix} &= \begin{bmatrix} \mathbf{0} & -\mathbf{K}_{21}\mathbf{K}_{11}^{-1}\mathbf{K}_{12} + \mathbf{K}_{22} \end{bmatrix} \begin{bmatrix} \boldsymbol{\theta}_0 \\ \boldsymbol{\theta}_C \end{bmatrix} + \\
&\quad (-\mathbf{K}_{21}\mathbf{K}_{11}^{-1}\mathbf{K}_{b1} + \mathbf{K}_{b2})\mathbf{b} + \begin{bmatrix} -\mathbf{K}_{21}\mathbf{K}_{11}^{-1} & \mathbf{I}_{22} \end{bmatrix} \begin{bmatrix} \mathbf{f}_0 \\ \mathbf{f}_C \end{bmatrix}
\end{aligned} \tag{16}
$$

All coefficients of $\boldsymbol{\theta}_0$ in equation (16) are zero which implies that this equation does not depend on temperatures of nodes without thermal capacity,

$$\mathbf{C}_C \dot{\boldsymbol{\theta}}_C = (-\mathbf{K}_{21}\mathbf{K}_{11}^{-1}\mathbf{K}_{12} + \mathbf{K}_{22})\boldsymbol{\theta}_C + (-\mathbf{K}_{21}\mathbf{K}_{11}^{-1}\mathbf{K}_{b1} + \mathbf{K}_{b2})\mathbf{b} + \begin{bmatrix} -\mathbf{K}_{21}\mathbf{K}_{11}^{-1} & \mathbf{I}_{22} \end{bmatrix}\begin{bmatrix} \mathbf{f}_0 \\ \mathbf{f}_C \end{bmatrix} \quad (17)$$

Since the matrix $\mathbf{C}_C$ in equation (17) is invertible, the state space representation of the modeled thermal network is:

$$\dot{\boldsymbol{\theta}}_C = \mathbf{C}_C^{-1}(-\mathbf{K}_{21}\mathbf{K}_{11}^{-1}\mathbf{K}_{12} + \mathbf{K}_{22})\boldsymbol{\theta}_C + \mathbf{C}_C^{-1}\begin{bmatrix} -\mathbf{K}_{21}\mathbf{K}_{11}^{-1}\mathbf{K}_{b1} + \mathbf{K}_{b2} & -\mathbf{K}_{21}\mathbf{K}_{11}^{-1} & \mathbf{I}_{22} \end{bmatrix}\begin{bmatrix} \mathbf{b} \\ \mathbf{f}_0 \\ \mathbf{f}_C \end{bmatrix} \quad (18)$$

or

$$\dot{\boldsymbol{\theta}}_C = \mathbf{A}_S \boldsymbol{\theta}_C + \mathbf{B}_S \mathbf{u} \quad (19)$$

where the state matrix is

$$\mathbf{A}_S = \mathbf{C}_C^{-1}(-\mathbf{K}_{21}\mathbf{K}_{11}^{-1}\mathbf{K}_{12} + \mathbf{K}_{22}) \quad (20)$$

the input matrix is

$$\mathbf{B}_S = \mathbf{C}_C^{-1}\begin{bmatrix} -\mathbf{K}_{21}\mathbf{K}_{11}^{-1}\mathbf{K}_{b1} + \mathbf{K}_{b2} & -\mathbf{K}_{21}\mathbf{K}_{11}^{-1} & \mathbf{I}_{22} \end{bmatrix} \quad (21)$$

and the input vector is:

$$\mathbf{u} = \begin{bmatrix} \mathbf{b} & \mathbf{f}_0 & \mathbf{f}_C \end{bmatrix}^T \quad (22)$$

Considering that the output is $\theta_a$, the state space model is:

$$\begin{aligned} \dot{\boldsymbol{\theta}}_C &= \mathbf{A}_S \, \boldsymbol{\theta}_C + \mathbf{B}_S \, \mathbf{u} \\ \theta_a &= \mathbf{C}_S \, \boldsymbol{\theta}_C + \mathbf{D}_S \, \mathbf{u} \end{aligned} \quad (23)$$

If the node $\theta_a$ has a thermal capacity (i.e. if $\theta_a$ is a state variable), the matrix $\mathbf{C}_S$ extracts $\theta_a$ from the vector $\boldsymbol{\theta}_C$ and the matrix $\mathbf{D}_S$ is zero. If the node $\theta_a$ does not have a thermal capacity (i.e. if $\theta_a$ is not a state variable), then $\theta_a$ is an element of the vector $\boldsymbol{\theta}_0$. The vector of temperatures $\boldsymbol{\theta}_0$ can be obtained from the first row of the equation (15):

$$\boldsymbol{\theta}_0 = -\mathbf{K}_{11}^{-1}\left(\mathbf{K}_{12}\boldsymbol{\theta}_C + \mathbf{K}_{b1}\mathbf{b} + \mathbf{I}_{11}\mathbf{f}_0\right) = -\mathbf{K}_{11}^{-1}\left(\mathbf{K}_{12}\boldsymbol{\theta}_C + \begin{bmatrix} \mathbf{K}_{b1} & \mathbf{I}_{11} & \mathbf{0} \end{bmatrix}\begin{bmatrix} \mathbf{b} \\ \mathbf{f}_0 \\ \mathbf{f}_C \end{bmatrix}\right) \quad (24)$$

or

$$\boldsymbol{\theta}_0 = \mathbf{C}_S \boldsymbol{\theta}_C + \mathbf{D}_S \mathbf{u} \quad (25)$$

where

$$\mathbf{C}_S = -\mathbf{K}_{11}^{-1} \mathbf{K}_{12} \quad (26)$$

$$\mathbf{D}_s = -\mathbf{K}_{11}^{-1}[\mathbf{K}_{b1} \quad \mathbf{I}_{11} \quad \mathbf{0}] \tag{27}$$

and

$$\mathbf{u} = \begin{bmatrix} \mathbf{b} & \mathbf{f}_0 & \mathbf{f}_C \end{bmatrix}^T \tag{28}$$

Then $\theta_a$ can be extracted from the vector of temperatures $\mathbf{\theta}_0$.

## 5    Transfer function representation of the thermal model

The relation between the inputs, $\mathbf{u}$, and the output, $\theta_a$, can be expressed as a set of transfer functions or a transfer matrix. Applying Laplace transform for zero initial conditions, the set of equations (23) become:

$$\begin{aligned} s\mathbf{\theta}_C &= \mathbf{A}_S\mathbf{\theta}_C + \mathbf{B}_S\mathbf{u} \\ \theta_a &= \mathbf{C}_S\mathbf{\theta}_C + \mathbf{D}_S\mathbf{u} \end{aligned} \tag{29}$$

where $s = \sigma + j\omega$ is the complex variable. From the first equation of (29), we obtain $(s\mathbf{I} - \mathbf{A}_S)\mathbf{\theta}_C = \mathbf{B}_S\mathbf{u}$. Eliminating the state vector $\mathbf{\theta}_C$ from (29), we obtain

$$\theta_a = [\mathbf{C}_S(s\mathbf{I} - \mathbf{A}_S)^{-1}\mathbf{B}_S + \mathbf{D}_S]\mathbf{u} \tag{30}$$

where

$$\mathbf{H} = \mathbf{C}_S(s\mathbf{I} - \mathbf{A}_S)^{-1}\mathbf{B}_S + \mathbf{D}_S \tag{31}$$

is the transfer matrix in which each element is a transfer function relating the output $\theta_a$ to a particular input from the input vector $\mathbf{u}$. Conduction transfer functions (CFT) are widely used in thermal load calculation. The coefficients of the transfer functions can be calculated precisely (Li *et al.* 2009).

## 6    Transfer function model in the case of non-negligible thermal capacity of the indoor air

We will consider a thermal network (Figure 3 a) that is an epitome of the heat balance method (Figure 1). Since the aim is to present the principle, the model from Figure 3 keeps the essential aspects of the heat balance method at a minimum number of temperature nodes. Normally, the wall would consist of several layers, each layer discretized in order to obtain temperatures at several nodes inside it. Generally, different surfaces would be considered for the room walls, windows, doors, floors, ceilings, etc. Thermal bridges can be also included in the model by treating heat conduction in two and three dimensions. But the simplicity of the model shown in Figure 3 does not reduce its generality for the problem presented hereafter.

Any thermal network can be easily put to comply with the typical branch and node shown in Figure 2. For example, the thermal network of Figure 3b is the equivalent of the network represented in Figure 3a. For a numerical example, let us consider that the elements of the circuit shown in Figure 3 have the values given in Table 1.

For the thermal network presented in Figure 3, the temperature differences for each thermal resistance are:

$$\begin{cases} e_1 = T_o - \theta_a \\ e_2 = T_o - \theta_{so} \\ e_3 = \theta_{so} - \theta_w \\ e_4 = \theta_w - \theta_{si} \\ e_5 = \theta_{si} - \theta_a \end{cases} \tag{32}$$

or

$$\begin{bmatrix} e_1 \\ e_2 \\ e_3 \\ e_4 \\ e_5 \end{bmatrix} = - \begin{bmatrix} 0 & 0 & 1 & 0 \\ 1 & 0 & 0 & 0 \\ -1 & 0 & 0 & 1 \\ 0 & 1 & 0 & -1 \\ 0 & -1 & 1 & 0 \end{bmatrix} \begin{bmatrix} \theta_{so} \\ \theta_{si} \\ \theta_a \\ \theta_w \end{bmatrix} + \begin{bmatrix} T_{ov} \\ T_{ow} \\ 0 \\ 0 \\ 0 \end{bmatrix} \tag{33}$$

By noting

$$\mathbf{e} = \begin{bmatrix} e_1 & e_2 & e_3 & e_4 & e_5 \end{bmatrix}^T$$

$$\mathbf{b} = \begin{bmatrix} T_o & T_o & 0 & 0 & 0 \end{bmatrix}^T$$

$$\boldsymbol{\theta} = \begin{bmatrix} \theta_{so} & \theta_{si} & \theta_a & \theta_w \end{bmatrix}^T$$

and the incidence matrix:

$$\mathbf{A} = \begin{matrix} & \theta_{so} & \theta_{si} & \theta_a & \theta_w & \\ & \begin{bmatrix} 0 & 0 & 1 & 0 \\ 1 & 0 & 0 & 0 \\ -1 & 0 & 0 & 1 \\ 0 & 1 & 0 & -1 \\ 0 & -1 & 1 & 0 \end{bmatrix} & \begin{matrix} R_v \\ R_{co} \\ R_{w1} \\ R_{w2} \\ R_{ci} \end{matrix} \end{matrix} \tag{34}$$

the set of equations (32) or (33) may be written as:

$$\mathbf{e} = -\mathbf{A}\boldsymbol{\theta} + \mathbf{b} \tag{35}$$

The heat transfer rates in the network branches can be expressed as

$$\begin{cases} q_1 \equiv q_v = R_v^{-1} e_1 \\ q_2 \equiv q_{co} = R_{co}^{-1} e_2 \\ q_3 \equiv q_{w1} = R_{w1}^{-1} e_3 \\ q_4 \equiv q_{w2} = R_{w2}^{-1} e_4 \\ q_5 \equiv q_{ci} = R_{ci}^{-1} e_2 \end{cases} \tag{36}$$

or

$$\begin{bmatrix} q_1 \\ q_2 \\ q_3 \\ q_4 \\ q_5 \end{bmatrix} = \begin{bmatrix} R_v^{-1} & 0 & 0 & 0 & 0 \\ 0 & R_{co}^{-1} & 0 & 0 & 0 \\ 0 & 0 & R_{w1}^{-1} & 0 & 0 \\ 0 & 0 & 0 & R_{w2}^{-1} & 0 \\ 0 & 0 & 0 & 0 & R_{ci}^{-1} \end{bmatrix} \begin{bmatrix} e_1 \\ e_2 \\ e_3 \\ e_4 \\ e_5 \end{bmatrix} \tag{37}$$

By putting:

$$\mathbf{q} = \begin{bmatrix} q_1 & q_2 & q_3 & q_4 & q_5 \end{bmatrix}^T$$

and

$$\mathbf{G} = \begin{bmatrix} R_v^{-1} & 0 & 0 & 0 & 0 \\ 0 & R_{co}^{-1} & 0 & 0 & 0 \\ 0 & 0 & R_{w1}^{-1} & 0 & 0 \\ 0 & 0 & 0 & R_{w2}^{-1} & 0 \\ 0 & 0 & 0 & 0 & R_{ci}^{-1} \end{bmatrix}$$

*Table 1. Parameter of the thermal network shown in Figure 3*

| Parameter | Symbol in Figure 3 | Value |
|---|---|---|
| Walls capacity [J/K] | $C_w$ | $4 \cdot 10^6$ |
| Indoor air capacity [J/K] | $C_a$ | $82 \cdot 10^3$ |
| Thermal conductance of the walls [W/K] | $R_w^{-1}$ | 1.45 |
| Thermal conductance of half of the wall [W/K] | $R_{w1}^{-1} = R_{w2}^{-1}$ | 2.90 |
| Thermal conductance of the windows and due to losses by ventilation [W/K] | $R_v^{-1}$ | 38.3 |
| Outdoor convection conductance [W/K] | $R_{co}^{-1}$ | 250.0 |
| Indoor convection conductance [W/K] | $R_{ci}^{-1}$ | 125.0 |

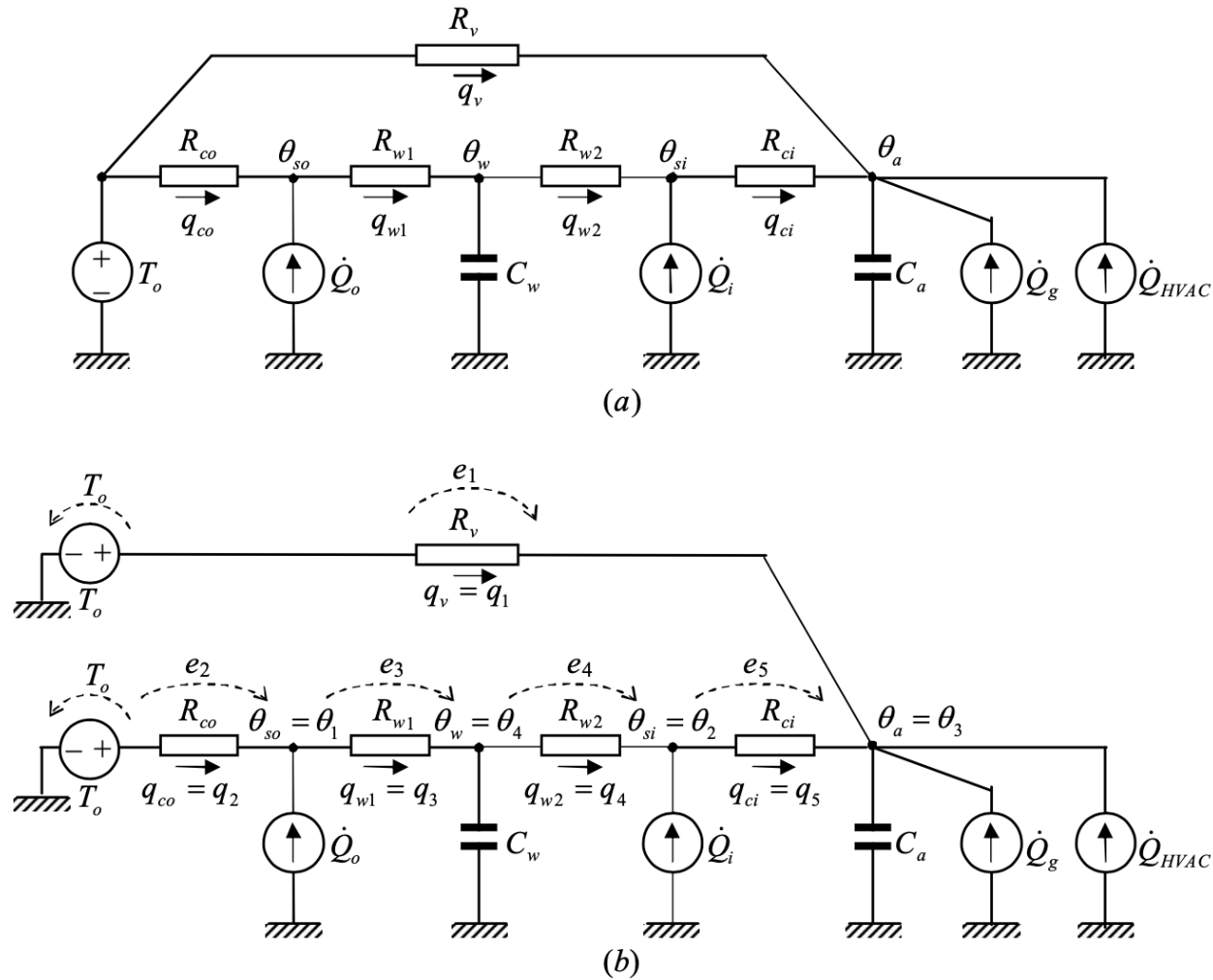

*Figure 3 Epitome thermal network for the heat balance method: a) usual representation; b) representation in the form of typical branches (according to Figure 2)*

the set of equations (36) or (37), representing the heat flow in each branch, can be written as

$$\mathbf{q} = \mathbf{Ge} \qquad (38)$$

The heat balance in each temperature node $\boldsymbol{\theta} = [\theta_{so} \quad \theta_{si} \quad \theta_a \quad \theta_w]^T$ gives:

$$\begin{cases} 0 = q_2 - q_3 + \dot{Q}_o \\ 0 = q_4 - q_5 + \dot{Q}_i \\ C_a \dot{\theta}_a = q_1 + q_5 + \dot{Q}_g + \dot{Q}_{HVAC} \\ C_w \dot{\theta}_w = q_3 - q_4 \end{cases} \qquad (39)$$

It is worth noting that the 3$^{rd}$ equation,

$C_a \dot{\theta}_a = q_1 + q_5 + \dot{Q}_g + \dot{Q}_{HVAC}$, where $q_1 \equiv q_v$ and $q_5 \equiv q_{ci}$

from the system of equations (39) is, in fact, the same as equation (1) used in the heat balance method. In matrix form, the system of equations (39) becomes

$$\begin{bmatrix} 0 & 0 & 0 & 0 \\ 0 & 0 & 0 & 0 \\ 0 & 0 & C_a & 0 \\ 0 & 0 & 0 & C_w \end{bmatrix} \begin{bmatrix} \theta_{so} \\ \theta_{si} \\ \dot{\theta}_a \\ \dot{\theta}_w \end{bmatrix} = \begin{bmatrix} 0 & 1 & -1 & 0 & 0 \\ 0 & 0 & 0 & 1 & -1 \\ 1 & 0 & 0 & 0 & 1 \\ 0 & 0 & 1 & -1 & 0 \end{bmatrix} \begin{bmatrix} q_1 \\ q_2 \\ q_3 \\ q_4 \\ q_5 \end{bmatrix} + \begin{bmatrix} \dot{Q}_o \\ \dot{Q}_i \\ \dot{Q}_g + \dot{Q}_{HVAC} \\ 0 \end{bmatrix} \qquad (40)$$

By putting

$$\mathbf{f} = [\dot{Q}_o \quad \dot{Q}_i \quad \dot{Q}_g + \dot{Q}_{HVAC} \quad 0]^T,$$

$$\mathbf{C} = \begin{bmatrix} 0 & 0 & 0 & 0 \\ 0 & 0 & 0 & 0 \\ 0 & 0 & C_a & 0 \\ 0 & 0 & 0 & C_w \end{bmatrix}$$

and observing that the matrix operator for differences of flows in temperature nodes is the transpose of the incidence matrix,

$$\mathbf{A}^T = \begin{bmatrix} 0 & 1 & -1 & 0 & 0 \\ 0 & 0 & 0 & 1 & -1 \\ 1 & 0 & 0 & 0 & 1 \\ 0 & 0 & 1 & -1 & 0 \end{bmatrix},$$

equation (40) becomes

$$\mathbf{C\dot{\theta}} = \mathbf{A}^T\mathbf{q} + \mathbf{f} \tag{41}$$

Substituting $\mathbf{q} = \mathbf{Ge}$ from equation (38) and $\mathbf{e} = -\mathbf{A\theta} + \mathbf{b}$ from equation (35), the equation (41) becomes

$$\mathbf{C\dot{\theta}} = -\mathbf{A}^T\mathbf{GA\theta} + \mathbf{A}^T\mathbf{Gb} + \mathbf{f} \tag{42}$$

or

$$\mathbf{C\dot{\theta}} = \mathbf{K\theta} + \mathbf{K}_b\mathbf{b} + \mathbf{f} \tag{43}$$

where

$$\mathbf{G} = \begin{bmatrix} 38.3 & 0 & 0 & 0 & 0 \\ 0 & 250.0 & 0 & 0 & 0 \\ 0 & 0 & 2.9 & 0 & 0 \\ 0 & 0 & 0 & 2.9 & 0 \\ 0 & 0 & 0 & 0 & 125 \end{bmatrix} \tag{44}$$

and

$$\mathbf{C} = \begin{bmatrix} 0 & 0 & 0 & 0 \\ 0 & 0 & 0 & 0 \\ 0 & 0 & 82\cdot10^3 & 0 \\ 0 & 0 & 0 & 4\cdot10^6 \end{bmatrix} \tag{45}$$

The partitioned matrix $\mathbf{K}$ is

$$\mathbf{K} = -\mathbf{A}^T\mathbf{GA} = \left[\begin{array}{cc:cc} -252.9 & 0 & 0 & 2.9 \\ 0 & -127.9 & 125.0 & 2.9 \\ \hdashline 0 & 125.0 & -163.3 & 0 \\ 2.9 & 2.9 & 0 & -5.8 \end{array}\right] \tag{46}$$

where the blocs are

$$\mathbf{K}_{11} = \begin{bmatrix} -252.9 & 0 \\ 0 & -127.9 \end{bmatrix}; \quad \mathbf{K}_{12} = \begin{bmatrix} 0 & 2.9 \\ 125.0 & 2.9 \end{bmatrix}$$

$$\mathbf{K}_{21} = \begin{bmatrix} 0 & 125.0 \\ 2.9 & 2.9 \end{bmatrix}; \quad \mathbf{K}_{22} = \begin{bmatrix} -163.3 & 0 \\ 0 & -5.8 \end{bmatrix}$$

and the partitioned matrix $\mathbf{K}_b$ is:

$$\mathbf{K}_b = \mathbf{A}^T \mathbf{G} = \begin{bmatrix} 0 & 250 & -2.9 & 0 & 0 \\ 0 & 0 & 0 & 2.9 & -125 \\ \hdashline 38.3 & 0 & 0 & 0 & 125 \\ 0 & 0 & 2.9 & -2.9 & 0 \end{bmatrix} \tag{47}$$

where the blocs are

$$\mathbf{K}_{b1} = \begin{bmatrix} 0 & 250 & -2.9 & 0 & 0 \\ 0 & 0 & 0 & 2.9 & -125 \end{bmatrix}$$

$$\mathbf{K}_{b2} = \begin{bmatrix} 38.3 & 0 & 0 & 0 & 125 \\ 0 & 0 & 2.9 & -2.9 & 0 \end{bmatrix}$$

The state-space model is

$$\dot{\boldsymbol{\theta}}_C = \mathbf{C}_C^{-1}(-\mathbf{K}_{21}\mathbf{K}_{11}^{-1}\mathbf{K}_{12} + \mathbf{K}_{22})\boldsymbol{\theta}_C + \mathbf{C}_C^{-1}\left[-\mathbf{K}_{21}\mathbf{K}_{11}^{-1}\mathbf{K}_{b1} + \mathbf{K}_{b2} \quad -\mathbf{K}_{21}\mathbf{K}_{11}^{-1} \quad \mathbf{I}_{22}\right]\begin{bmatrix} \mathbf{b} \\ \mathbf{f}_0 \\ \mathbf{f}_C \end{bmatrix} \tag{48}$$

where

$$\mathbf{C}_C = \begin{bmatrix} 82 \cdot 10^3 & 0 \\ 0 & 4 \cdot 10^6 \end{bmatrix}$$

and

$$\mathbf{I}_{22} = \begin{bmatrix} 1 & 0 \\ 0 & 1 \end{bmatrix}$$

Equation (48) may be written as:

$$\dot{\boldsymbol{\theta}}_C = \mathbf{A}_S \boldsymbol{\theta}_C + \mathbf{B}_S \mathbf{u} \tag{49}$$

where the state matrix is:

$$\mathbf{A}_S = \mathbf{C}_C^{-1}(-\mathbf{K}_{21}\mathbf{K}_{11}^{-1}\mathbf{K}_{12} + \mathbf{K}_{22}) = \begin{bmatrix} -0.501 \cdot 10^{-3} & 0.034 \cdot 10^{-3} \\ 0.708 \cdot 10^{-6} & -1.452 \cdot 10^{-6} \end{bmatrix} \tag{50}$$

and the input matrix is:

$$\mathbf{B}_S = \mathbf{C}_C^{-1}\left[-\mathbf{K}_{21}\mathbf{K}_{11}^{-1}\mathbf{K}_{b1} + \mathbf{K}_{b2} \quad -\mathbf{K}_{21}\mathbf{K}_{11}^{-1} \quad \mathbf{I}_{22}\right] \tag{51}$$

The input vector is

$$\mathbf{u} = \begin{bmatrix} \mathbf{b} & \mathbf{f}_0 & \mathbf{f}_C \end{bmatrix}^T = \begin{bmatrix} T_o & T_o & 0 & 0 & 0 & \dot{Q}_o & \dot{Q}_i & \dot{Q}_g + \dot{Q}_{HVAC} & 0 \end{bmatrix}^T \qquad (52)$$

where $\mathbf{b} = \begin{bmatrix} T_o & T_o & 0 & 0 & 0 \end{bmatrix}^T$, $\mathbf{f}_0 = \begin{bmatrix} \dot{Q}_0 & \dot{Q}_i \end{bmatrix}^T$ and $\mathbf{f}_C = \begin{bmatrix} \dot{Q}_g + \dot{Q}_{HVAC} & 0 \end{bmatrix}^T$. However, we can define the input vector by taking only the elements 1, 2, 6, 7, and 8 from the input vector $\mathbf{u}$ given by (52),

$$\mathbf{u} = \begin{bmatrix} T_{ov} & T_{ow} & \dot{Q}_o & \dot{Q}_i & \dot{Q}_g + \dot{Q}_{HVAC} \end{bmatrix}^T \qquad (53)$$

and, therefore, only the columns 1, 2, 6, 7, and 8 from the matrix $\mathbf{B}_S$,

$$\mathbf{B}_S = \begin{bmatrix} 4.7 \cdot 10^{-4} & 0 & 0 & 1.2 \cdot 10^{-5} & 1.2 \cdot 10^{-5} \\ 0 & 7.2 \cdot 10^{-7} & 2.9 \cdot 10^{-9} & 5.7 \cdot 10^{-9} & 0 \end{bmatrix} \qquad (54)$$

Putting

$$\mathbf{C}_S = \begin{bmatrix} 1 & 0 \end{bmatrix} \qquad (55)$$

the state space model corresponding to the thermal circuit from Figure 3 with the parameters given in Table 1 is

$$\begin{aligned} \dot{\boldsymbol{\theta}}_C &= \mathbf{A}_S \boldsymbol{\theta}_C + \mathbf{B}_S \mathbf{u} \\ \theta_a &= \mathbf{C}_S \boldsymbol{\theta}_C \end{aligned} \qquad (56)$$

where $\boldsymbol{\theta}_C = \begin{bmatrix} \theta_a & \theta_w \end{bmatrix}^T$, $\mathbf{u}$ is given by (53), $\mathbf{A}_S$ is given by (50), $\mathbf{B}_S$ is given by (54), and $\mathbf{C}_S$ is given by (55).

The transfer matrix corresponding to the state space model (56) is obtained by substituting $\mathbf{A}_S$ $\mathbf{B}_S$ and $\mathbf{C}_S$ in equation (31):

$$\mathbf{H} = \begin{bmatrix} H_1 & H_2 & H_3 & H_4 & H_5 \end{bmatrix} = \begin{bmatrix} \dfrac{6.764 \cdot 10^5 s + 9.641 \cdot 10^{-1}}{1.448 \cdot 10^9 s^2 + 7.286 \cdot 10^5 s + 1} \\[2ex] \dfrac{3.587 \cdot 10^{-2}}{1.448 \cdot 10^9 s^2 + 7.286 \cdot 10^5 s + 1} \\[2ex] \dfrac{1.435 \cdot 10^{-4}}{1.448 \cdot 10^9 s^2 + 7.286 \cdot 10^5 s + 1} \\[2ex] \dfrac{1.726 \cdot 10^4 s + 2.488 \cdot 10^{-2}}{1.448 \cdot 10^9 s^2 + 7.286 \cdot 10^5 s + 1} \\[2ex] \dfrac{1.766 \cdot 10^4 s + 2.517 \cdot 10^{-2}}{1.448 \cdot 10^9 s^2 + 7.286 \cdot 10^5 s + 1} \end{bmatrix}^T$$

By using the transfer functions from the transfer matrix, the output is

$$\theta_a = \mathbf{H}\,\mathbf{u} = \begin{bmatrix} H_1 & H_2 & H_3 & H_4 & H_5 \end{bmatrix} \begin{bmatrix} T_{ov} \\ T_{ow} \\ \dot{Q}_o \\ \dot{Q}_i \\ \dot{Q}_g + \dot{Q}_{HVAC} \end{bmatrix}$$

or

$$\theta_a = H_1 T_o + H_2 T_o + H_3 \dot{Q}_o + H_4 \dot{Q}_i + H_5 (\dot{Q}_g + \dot{Q}_{HVAC}) \tag{57}$$

All transfer functions $H_1$, …, $H_5$ are strictly proper (they respect the physical causality).

It is worth noting that the heat balance on air volume, expressed by equation (1) as the 3rd equation of the system of equations (39), was used to obtain the expression (57). However, this is not evident at first sight.

## 7 Transfer function model in the case of negligible thermal capacity of the indoor air

In the case of negligible capacity in the node $\theta_a$, i.e. $C_a = 0$, the partitioned matrix $\mathbf{K}$ from equation (46) becomes

$$\mathbf{K} = -\mathbf{A}^T \mathbf{G}\mathbf{A} = \left[ \begin{array}{ccc:c} -252.9 & 0 & 0 & 2.9 \\ 0 & -127.9 & 125.0 & 2.9 \\ 0 & 125.0 & -163.3 & 0 \\ \hdashline 2.9 & 2.9 & 0 & -5.8 \end{array} \right] = \begin{bmatrix} \mathbf{K}_{11} & \mathbf{K}_{12} \\ \mathbf{K}_{21} & \mathbf{K}_{22} \end{bmatrix}$$

and the partitioned matrix $\mathbf{K}_b$ from equation (47) becomes

$$\mathbf{K}_b = \mathbf{A}^T\mathbf{G} = \begin{bmatrix} 0 & 250 & -2.9 & 0 & 0 \\ 0 & 0 & 0 & 2.9 & -125 \\ 38.3 & 0 & 0 & 0 & 125 \\ \hdashline 0 & 0 & 2.9 & -2.9 & 0 \end{bmatrix} = \begin{bmatrix} \mathbf{K}_{b1} \\ \mathbf{K}_{b2} \end{bmatrix}$$

We obtain:

$$\mathbf{A}_S = \mathbf{C}_C^{-1}(-\mathbf{K}_{21}\mathbf{K}_{11}^{-1}\mathbf{K}_{12} + \mathbf{K}_{22}) = -1.376 \cdot 10^{-6} \tag{58}$$

$$\mathbf{B}_S = \begin{bmatrix} 6.597 \cdot 10^{-7} & 7.167 \cdot 10^{-7} & 2.867 \cdot 10^{-7} & 2.250 \cdot 10^{-7} & 1.723 \cdot 10^{-7} \end{bmatrix} \tag{59}$$

$$\mathbf{C}_S = 6.890 \cdot 10^{-2} \tag{60}$$

and, by using equation (27),

$$\mathbf{D}_S = \begin{bmatrix} 0.9311 & 0 & 0 & 0.023759 & 0.024311 \end{bmatrix}^T \tag{61}$$

Substituting in equation (31) the values of the matrices $\mathbf{A}_S, \mathbf{B}_S, \mathbf{C}_S$ and $\mathbf{D}_S$ given above, we obtain the matrix transfer:

$$\mathbf{H} = \begin{bmatrix} H_1 & H_2 & H_3 & H_4 & H_5 \end{bmatrix} = \begin{bmatrix} \dfrac{6.764 \cdot 10^{-2}s + 9.641 \cdot 10^{-1}}{7.265 \cdot 10^5 s + 1} \\[3mm] \dfrac{3.587 \cdot 10^{-2}}{7.265 \cdot 10^5 s + 1} \\[3mm] \dfrac{1.435 \cdot 10^{-4}}{7.265 \cdot 10^5 s + 1} \\[3mm] \dfrac{1.726 \cdot 10^{-4}s + 2.488 \cdot 10^{-2}}{7.265 \cdot 10^5 s + 1} \\[3mm] \dfrac{1.766 \cdot 10^{-4}s + 2.517 \cdot 10^{-3}}{7.265 \cdot 10^5 s + 1} \end{bmatrix}^T$$

As for the case of non-negligible thermal capacity of the indoor air, all transfer functions $H_1$, ..., $H_5$ are proper or strictly proper. By using the transfer functions, the output is

$$\theta_a = \mathbf{H}\,\mathbf{u} = \begin{bmatrix} H_1 & H_2 & H_3 & H_4 & H_5 \end{bmatrix} \begin{bmatrix} T_{ov} \\ T_{ow} \\ \dot{Q}_o \\ \dot{Q}_i \\ \dot{Q}_g + \dot{Q}_{HVAC} \end{bmatrix}$$

or

$$\theta_a = H_1 T_o + H_2 T_o + H_3 \dot{Q}_o + H_4 \dot{Q}_i + H_5(\dot{Q}_g + \dot{Q}_{HVAC}) \tag{62}$$

As in the case of non-negligible thermal capacity of the room air, the heat balance equation (2), which is one of the equations from the system of equations (39), was used to obtain the output $\theta_a$. However, the relation between equation (2) and equation (62) is not self-evident.

## 8    Transfer function for the thermal load

In order to obtain the thermal load, the current method uses equation (3) or (4). By using the transfer functions, this is equivalent to express $\dot{Q}_{HVAC}$ from equation (57) or (62):

$$\dot{Q}_{HVAC} = H_5^{-1}\theta_a - H_5^{-1}H_1T_o - H_5^{-1}H_2T_o - H_5^{-1}H_3\dot{Q}_o - H_5^{-1}H_4\dot{Q}_i - \dot{Q}_g \qquad (63)$$

The transfer function $H_5^{-1}$, corresponding to input $\theta_a$, is improper. The transfer functions $H_5^{-1}H_1$ corresponding to the influence of the outdoor temperature, $T_o$, through ventilation and $H_5^{-1}H_4$ corresponding to the influence of the heat flux on the indoor walls, $\dot{Q}_i$, are proper. They correspond to inputs connected directly to the node $\theta_a$, i.e. not through a node containing a thermal capacity. The functions $H_5^{-1}H_2$ corresponding to the influence of the outdoor temperature, $T_o$, through walls and $H_5^{-1}H_3$ corresponding to influence of the heat flux on the outdoor wall, $\dot{Q}_o$, are strictly proper. They correspond to inputs connected to the node $\theta_a$ through nodes containing at least a thermal capacity.

It results that this method yields an improper function for input $\theta_a$. This is true not only for the two epitomes of non-negligible and negligible indoor air capacity presented above, but for any model of this type. For any thermal network constructed according to the heat balance method (Figure 1):
- the transfer function relating the load to the temperature in node $\theta_a$ will be always improper,
- the transfer functions relating the inputs to the output $\theta_a$ through branches which do not contain at least a thermal capacity will be strictly proper and
- the transfer functions connecting the inputs to the output $\theta_a$ by branches containing at least a thermal capacity will be proper.

## 9    Consequences of the improper transfer function

The models of multiple input – multiple output systems (MIMO) describe a causal relation from input to the output if and only if the transfer matrix is proper, i.e. if in any entry of the transfer matrix the degree of the denominator is larger than or equal to the degree of the numerator (Mareels and Polderman 1996). A system with an improper transfer matrix is not physically realizable because it requires prediction (Morari and Zafirou 1989). Furthermore, an improper transfer function amplifies the high frequencies.

Heat balance method for load calculation results in an improper transfer function. The practical consequence is that this method cannot be used to calculate the load, $\dot{Q}_{HVAC}$, when the space temperature, $\theta_a$, varies. For example, if the air space temperature has a step change (which is frequently the case in simulations), the high frequencies present in the signal of $\theta_a$ will be amplified. If the simulation time step tends to zero, the calculated load will tend to infinity, which technologically is obviously not achievable. One solution is to limit the value of the load (CEN 2007b), which implies that the simulation will not be used to calculate the maximum load; however, it can be useful for energy consumption estimation. Another solution is to introduce a term in the transfer function of $\theta_a$ in order to make it proper. This is achievable by considering the thermal inertia of the HVAC system or by introducing a feed-back control loop, in which case the maximum load depends on the controller type and settings (Ghiaus and Hazyuk 2010).

## 10  Conclusions

Heat balance method used for simulation of the time variation of the air space temperature as a function of inputs has strong physical basis. The most restrictive assumption is that the air in the thermal zone has a uniform temperature. The heat balance results in a set of differential algebraic equations which, after piece-wise linearization, can be written in state space form and then transformed in a set of transfer functions. These models respect the causality: the outputs depend on inputs through the state variables. Consequently, the transfer functions are proper.

Thermal load calculation is an inverse problem. One input of the physical system, the thermal load, becomes an output of the model and one output of the physical system, the indoor temperature, becomes an input of the model used in calculation. The method used to solve this problem does not respect the causality and results in an improper transfer function. Therefore, the high frequencies present in the signal of the indoor temperature will be amplified. The practical consequence is that, if the indoor temperature varies, the calculated load will depend on the simulation time step: smaller the time step, larger the calculated load. If the room temperature has a step variation, the calculated load will tend to infinity when the simulation step tends to zero. This effect is less evident if the simulation time step is one hour (the usual value in building simulation software) but becomes evident if the time step is reduced.

The problem of non causality can be solved by introducing a filter which will make the transfer function proper. This can be achieved by transforming the load calculation into a control problem. Then, the load will be the output of the control algorithm.

The issue of causality is not self-evident for complex systems. Therefore, the claim of simulation systems which are equation-based (i.e. in which the equations describe equalities, not assignments) that there is a benefit in the ability to choose arbitrarily the inputs and the outputs of the model (Modelica 2000, Dymola 2004) is questionable if this choice results in improper transfer functions.

**Acknowledgements**
This work was conducted in the framework of the International Energy Agency Annex 53 *Total energy use in buildings: analysis and evaluation methods* and in the research project *Fault detection and diagnosis of energy systems* (AIDE-3D) financed by the French National

Research Agency (ANR). Special thanks are given to Ion Hazyuk (INSA Lyon, France), Eric Blanco (Ecole Centrale de Lyon, France), Ad van der Aa (Cauherg-Huygen Raadgevende Ingenieurs BV, The Netherlands) and Ian Hensen (Eindhoven University of Technology, The Netherlands) for their interest, discussions, and support.

**References**

1. Bellenger, L., Bruning S., Pedersen, C., Romine, T., Wilkins C. (2001). Nonresidential cooling and heating load calculation procedures, in ASHRAE Handbook: Fundamentals, ISBN-10: 1931862516, USA.

2. CEN (2003). Heating systems in buildings — Method for calculation of the design heat load EN 12831:2003, European Committee for Standardization, Brussels

3. CEN (2007a). Thermal performance of buildings – Calculation of energy use for space heating and cooling – General criteria and validation procedures (Energy need calculations - heating and cooling) EN 15265:2007, European Committee for Standardization, Brussels

4. CEN (2007b). Thermal performance of buildings – Sensible room cooling load calculation – General criteria and validation procedures EN 15255:2007, European Committee for Standardization, Brussels

5. Clarke, J. (2001). *Energy Simulation in Building Design 2$^{nd}$ edition*. ISBN-10: 0750650826, Butterworth-Heinemann

6. Degiovanni, A. (1999). Transmission de l'énergie thermique – Conduction. *Techniques de l'Ingénieur*.

7. Dymola (2004). *Dymola. Dynamic Modeling Laboratory. User's Manual v 5.3a*. Dynasim AB, Sweden

8. EnergyPlus® (2012). *EnergyPlus Engineering Reference*. University of Illinois and Lawrence Berkeley National Laboratory

9. Ghiaus C., Hazyuk I. (2010). Calculation of optimal thermal load of intermittently heated buildings. Energy and Buildings, 42:1248-1258

10. Hernandez M., Medina M.A., Schruben D.L. (2003). Verification of an Energy Balance Approach to Estimate Indoor Wall Heat Fluxes Using Transfer Functions and Simplified Solar Heat Gain Calculations. Mathematical and Computer Modelling, 37: 235-243

11. Li X. Q., Chen Y., Spitler J.D., Fisher D. (2009). *Applicability of calculation methods for conduction transfer function of building constructions.* International Journal of Thermal Sciences, 48: 1441-1451

12. Mareels I., Polderman J. W. (1996). Adaptive Systems : An Introduction, Birkhäuser, Boston

13. Modelica (2000). *Modelica$^{TM}$ – A Unified Object-Oriented Language for Physical Systems Modeling*, v. 1.4, Modelica Association, Linköping, Sweden

14. Morari M., Zafirou E. (1989). Robust Process Control. Pretince-Hall Inc. ISBN 0-13-782153-0

15. Romagnoli J., Palazoglu A. (2012). Introduction to Process control 2nd edition, CRC Press, Taylor and Francis ISBN 978-1-4398-5486-0

16. Skeiker K. (2010). *Advanced software tool for the dynamic analysis of heat transfer in buildings: applications to Syria*. Energy, 35:2603-2609

17. Strang, G. (1986). *Introduction to Applied Mathematics*. Welley-Cambridge Press

18. Strang, G. (2009). *Introduction to Linear Algebra.* Welley-Cambridge Press
19. TRNSYS 16 (2007). *Mathematical Reference.* Solar Energy Laboratory, University of Wisconsin-Madison, WI, USA
20. Wang S., Chen Y. (2000). A novel and simple building load calculation model for building and system dynamical simulation. Applied Thermal Engineering, 21: 638-702
21. Xu X., Wang S. (2008). A simplified dynamic model for existing buildings using CTF and thermal network models. International Journal of Thermal Sciences 47: 1249-1262